\documentclass[10pt,aps,prc,floatfix,twocolumn,nofootinbib, superscriptaddress,preprintnumbers]{revtex4-1}

\usepackage{amsfonts,amsmath,amssymb,bm, xspace}
\usepackage{color,graphicx}
\usepackage{bbm}
\usepackage{dcolumn}                 
\graphicspath{{./figures/}}

\usepackage{hyperref}                

\usepackage{array,physics}
\usepackage{dcolumn}
\usepackage[normalem]{ulem}
\newcolumntype{d}[1]{D{.}{.}{#1}}

\newcommand{\Msol}{\,\text{M}_{\odot}\xspace}

\newcommand{\fmiq}{\, \text{fm}^{-3}}

\newcommand{\MeV}{\, \text{MeV}}

\newcommand{\sat}{\textrm{sat}}
\newcommand{\nsat}{n_\sat}

\newcommand{\tov}{\textrm{TOV}}

\hypersetup{%
    pdfsubject=Paper,
    pdfkeywords={nuclear physics} 
    unicode=true,
    breaklinks=true,
     colorlinks   = true,
     linkcolor = blue,
     citecolor = blue,
     menucolor = blue,
     urlcolor = blue
}

\begin{document}
\preprint{LA-UR-22-22575}

\title{Perturbative QCD and the Neutron Star Equation of State}

\author{R. Somasundaram}
\email{rsomasun@syr.edu}
\affiliation{Univ Lyon, Univ Claude Bernard Lyon 1, CNRS/IN2P3, IP2I Lyon, UMR 5822, F-69622, Villeurbanne, France}

\author{I. Tews}
\affiliation{Theoretical Division, Los Alamos National Laboratory, Los Alamos, New Mexico 87545, USA}

\author{J. Margueron}
\affiliation{Univ Lyon, Univ Claude Bernard Lyon 1, CNRS/IN2P3, IP2I Lyon, UMR 5822, F-69622, Villeurbanne, France}

\date{\today}

\begin{abstract}
We construct a physics-agnostic approach to the neutron star (NS) equation of state (EoS) based on a sound speed model, which connects both low-density information from nuclear theory and high-density constraints from perturbative QCD (pQCD).
Using this approach, we study the impact of pQCD calculations on NS EoS that have been constrained by astrophysical observations. 
We find that pQCD affects the EoS mainly beyond the densities realized in NS.
Furthermore, we observe an interesting interplay between pQCD and astrophysical constraints, with pQCD preferring softer EoS for the heaviest NS while recent NICER observations suggest the EoS to be stiffer.
We explore the sensitivity of our findings to pQCD uncertainties and study the constraining power of pQCD if future observations of heavy NS were to suggest radii larger than 13~km.
\end{abstract}

\maketitle

\textit{Introduction -}
Obtaining a consistent description of the Equation of State (EoS) of dense matter from low densities, where nuclear effective field theories (EFTs) are valid, $n \lesssim 2\nsat,$~\cite{Tews:2018kmu,Drischler:2020hwi} with $\nsat \approx 0.16 \fmiq$ being the nuclear saturation density, up to the highest densities explored in the universe, $n \approx 8 \nsat$, remains one of the major goals in nuclear physics~\cite{Lattimer:2004}. 
At lower densities $ n \lesssim 2\nsat$, advances in Chiral EFT ($\chi$EFT)~\cite{Epelbaum:2008ga,Machleidt:2011zz} allow for a microscopic description of nuclear matter consistent with the symmetries of Quantum Chromodynamics (QCD). 
Furthermore, properties of matter at these densities can be investigated in experiments with atomic nuclei~\cite{Tsang:2012se,Margueron2018a, Reed:2021nqk,Essick:2021kjb,Huth:2021bsp,Most:2022wgo}, allowing to calibrate high-precision Energy Density Functionals (EDF) that accurately reproduce such properties, see for instance Ref.~\cite{Bender2003} and references therein. 
At larger densities, the situation changes drastically. 
In this regime, effective nuclear potentials based on $\chi$EFT are no longer applicable due to the breakdown of the EFT~\cite{Tews:2018kmu,Drischler:2020hwi}. 
Furthermore, the extrapolation of EDFs well beyond the density regime where they were fit comes with systematic uncertainties that are difficult to quantify~\cite{Margueron2018a, Margueron2018b, Margueron2019,Somasundaram2021,Somasundaram:2021vgi}. Therefore, at these densities, our understanding of dense neutron-rich matter comes mainly from observations of neutron stars (NS), which are arguably one of the most fascinating objects in the universe, containing dense matter up to $n \equiv n_c^\tov \lesssim 8 \nsat$ in their cores~\cite{Rezzolla2018}. 
Recent multimessenger observations of NS, i.e., radio~\cite{Demorest:2010,Antoniadis:2013pzd,Cromartie:2019,Fonseca:2021wxt}, X-ray~\cite{Miller:2019,Riley:2019,Miller:2021qha,Riley:2021pdl}, gravitational-wave (GW) observations~\cite{TheLIGOScientific:2017,Abbott:2018exr, De:2018uhw}, and their electromagnetic (EM) counterparts~\cite{LIGOScientific:2017a,LIGOScientific:2017b}, have provided valuable new insights into the EoS of dense matter~\cite{Bauswein:2017vtn, Ruiz:2017due, Annala:2017llu, Radice:2017lry, Most:2018hfd, Radice:2018ozg, Lim:2018bkq, Tews:2018, Coughlin:2018fis, Capano:2019eae,  Dietrich:2020lps, Greif:2020pju, Guven:2020, Legred:2021hdx,Raaijmakers:2021uju, Huth:2021bsp, Essick:2021kjb}. 
Nevertheless, many open questions remain, such as if phase transitions occur in NS cores. 
Answering these questions requires both new experimental and observational data~\cite{Bauswein2019,Bauswein2020, Cierniak2021,Somasundaram:2021clp,Somasundaram2022}, as well as new theoretical constraints, such as the ones we discuss in this study.

A potential additional source of information can be added at asymptotically large densities~\cite{Kurkela:2014vha,Annala:2017llu,Annala:2019}, $n \equiv n_{\mathrm{pQCD}} \approx 40 \nsat$, where the theory governing the strong force (QCD) is perturbative, enabling calculations of the EoS of weakly interacting quark matter via a perturbative treatment of the QCD Lagrangian~\cite{Gorda:2018gpy,Gorda:2021znl}. 
For example, Ref.~\cite{Annala:2019} claimed to have found evidence for the presence of quark matter in NS cores by employing, among other constraints, perturbative QCD (pQCD) calculations at large densities.
While intriguing, pQCD calculations are valid above $n_{\mathrm{pQCD}}$, whereas stable NS are not expected to explore densities larger than $n_c^\tov$. 
Given the order of magnitude that separates these two density regimes, the importance of pQCD calculations in analyses of NS matter cannot be estimated straightforwardly.
Additional systematic uncertainties might be introduced by the choice of interpolation scheme between these different density regimes.

Komoltsev and Kurkela have recently suggested a new method to link both density regimes, allowing them to `integrate backwards', i.e., to propagate the pQCD constraints to lower densities in a completely general, analytical, and model-agnostic manner using only the thermodynamic potential and the conditions of causality and mechanical stability~\cite{Komoltsev:2021jzg}. 
They concluded that, neglecting NS observations, pQCD calculations exclude about 65\% of the area in the pressure-energy density plane at $n = 5 \nsat$. 
Here, we address whether pQCD constraints affect the NS EoS when existing constraints from astrophysical observations are accounted for, by employing a very general approach to extrapolate the EoS~\cite{Somasundaram:2021clp} to higher densities.

\textit{Approach -}
To reliably analyze the impact of the pQCD at asymptotically high densities and compare it to astrophysical and experimental constraints at lower densities, we rely here on the computational setup presented in Ref.~\cite{Somasundaram:2021clp}. 
This formalism is general enough to (i) capture our knowledge of the low-density EoS where nuclear physics constraints exist, (ii) fully explore the present uncertainties in observational NS data, and (iii) allow for the implementation of pQCD constraints minimizing the effects of uncontrolled interpolations over vastly separated density regions. 
For simplicity, the EoS up to $\nsat$ is given by the Douchin-Haensel SLY model~\cite{DH2001} based on the Skyrme SLy4 EDF~\cite{Chabanat:1997} that is well calibrated to the properties of nuclear matter and finite nuclei and commonly used in astrophysical applications. 
Beyond $\nsat$, we describe the EoS using an extension in the speed of sound plane, see Ref.~\cite{Somasundaram:2021clp} for more details. 
For each EoS, we then calculate the NS mass-radius-tidal deformability relations and analyze astrophysical observations of NS.

\begin{figure}
    \centering
    \includegraphics[scale=0.55]{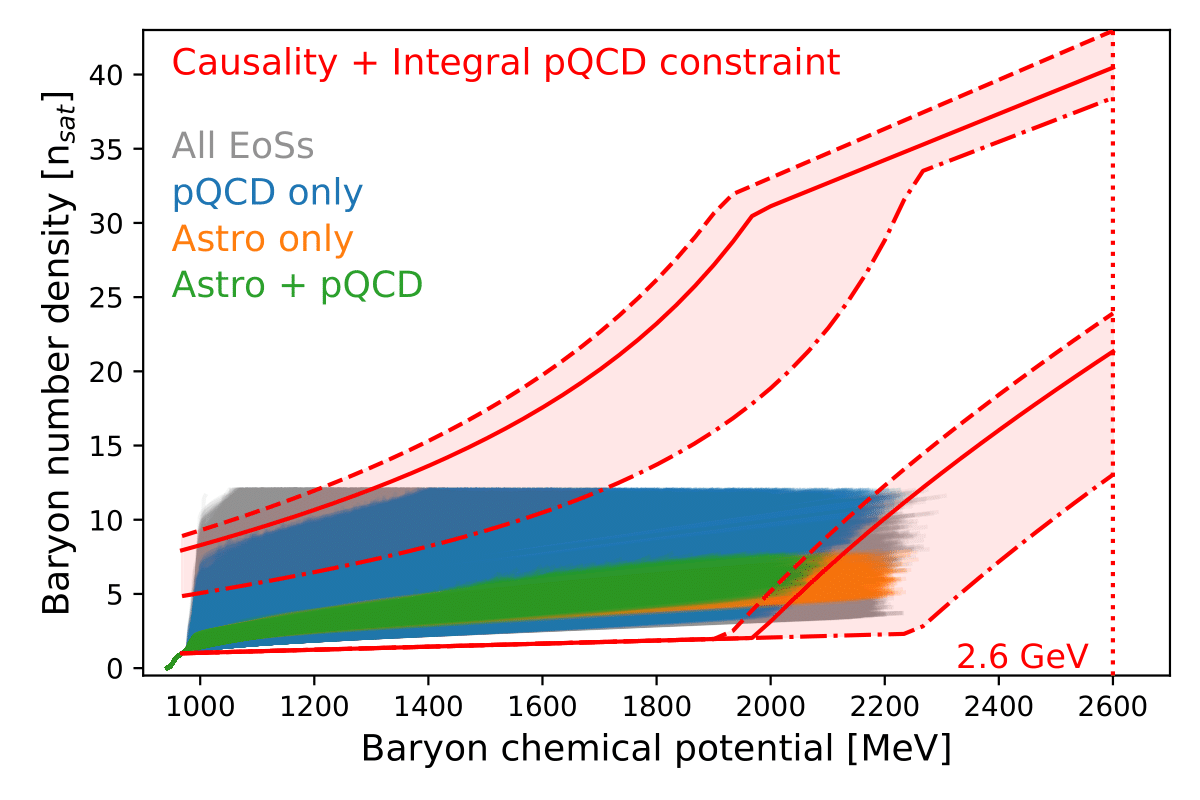}
    \includegraphics[scale=0.55]{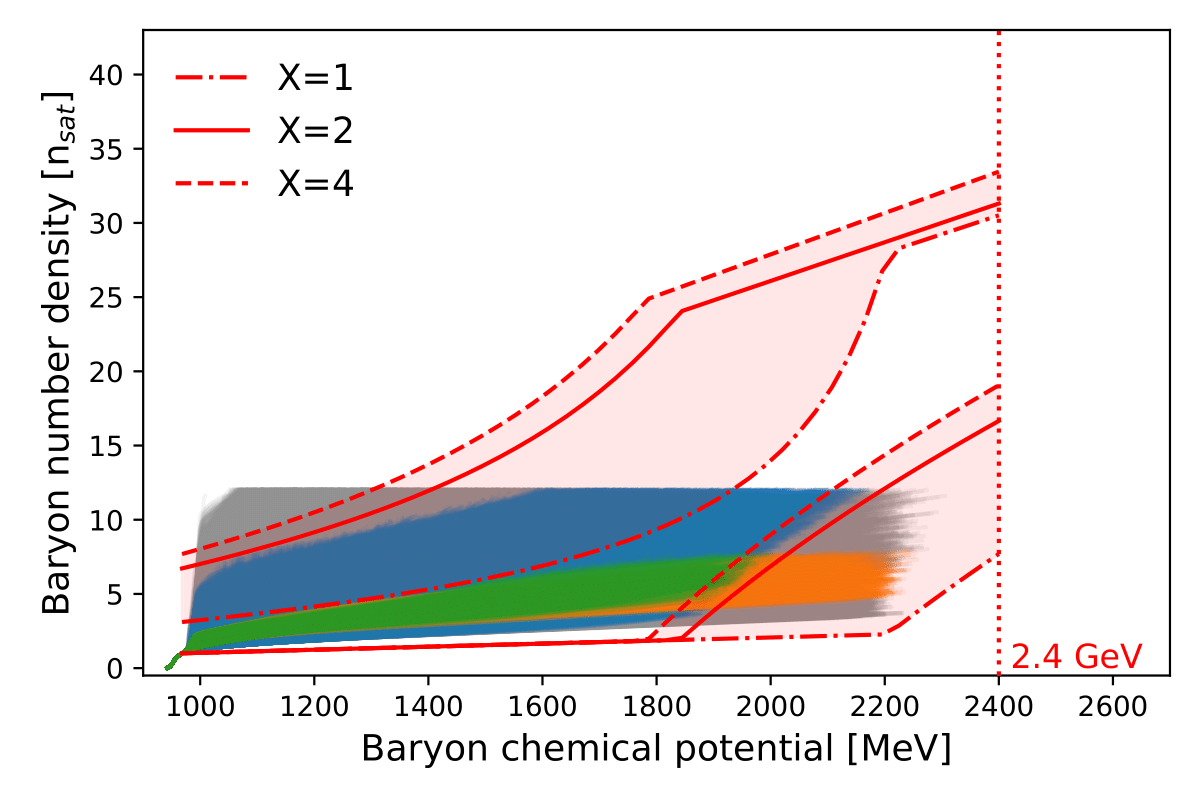}
    \caption{The baryon number density as function of the chemical potential. The red boundaries depict integrated constraints from pQCD with the red band showing the uncertainty (see text). 
    The top (bottom) panel is constructed using the pQCD EoS down to $\mu_{\mathrm{pQCD}} = 2.6$~GeV ($\mu_{\mathrm{pQCD}} = 2.4$~GeV).
    We show all EoS in our set (gray), the EoS constrained by pQCD, at $X=2$, only (blue), by astrophysical constraints only (orange), and by both astrophysical and pQCD constraints, at $X=2$, (green).}
    \label{fig:n_mu}
\end{figure}

We consider (i) the tidal deformability $\tilde{\Lambda}=222^{+420}_{-138}$ at 90\% confidence level (CL) for GW170817~\cite{De:2018}, which is consistent with other analyses~\cite{TheLIGOScientific:2017,Abbott:2018exr}, (ii) the independent analyses of X-ray observations of pulsars J0740+6620 and J0030+0451 by the NICER telescope~\cite{Miller:2019,Miller:2021qha,Riley:2019,Riley:2021pdl} by averaging over different results for the same source and (iii) radio observations of heavy NS imposing M$_\tov\gtrsim 2\Msol$~\cite{Cromartie:2019,Demorest:2010,Antoniadis:2013pzd}. 

In order to implement pQCD constraints, we follow the approach of Ref.~\cite{Komoltsev:2021jzg} by first imposing that $n_\textrm{min}(\mu) < n(\mu) < n_\textrm{max}(\mu)$ for $\mu \leq \mu_c^{\mathrm{TOV}}$, where $\mu_c^{\mathrm{TOV}}$ is the central chemical potential at the TOV limit and
\begin{equation}
  n_\textrm{max}(\mu) =
  \begin{cases}
   \frac{\mu^3 n_L - \mu \mu_L (\mu_L n_L + 2\Delta p)}{(\mu^2 + \mu_H^2)\mu_L} & \mu_L \leq \mu < \mu_c \\
    n_H\mu/\mu_H  & \mu_c \leq \mu \leq \mu_H
  \end{cases}
  \label{eq:n_max}
\end{equation}
and
\begin{equation}
  n_\textrm{min}(\mu) =
  \begin{cases}
   n_L\mu/\mu_L  & \mu_L \leq \mu \leq \mu_c \\
     \frac{\mu^3 n_H - \mu \mu_H (\mu_H n_H - 2\Delta p)}{(\mu^2 - \mu_L^2)\mu_H}  & \mu_c < \mu \leq \mu_H
  \end{cases}
  \label{eq:n_min}
\end{equation}
where $\mu_L$ is the chemical potential of the low-density Skyrme EoS at $n_L = n_\textrm{sat}$ and $n_H$, $\mu_H$ are the corresponding pQCD values evaluated at $\mu_H = \mu_\textrm{pQCD}$. Also, $\Delta p = p_H - p_L$ and $\mu_c$ is given by the intercept of the causal line and the integral constraint, see Ref.~\cite{Komoltsev:2021jzg}.  

In this work, we use the partial N3LO calculation of Ref.~\cite{Gorda:2021znl} for the high density pQCD EoS which determines the thermodynamic variables $p_H (\mu_H)$ and $n_H (\mu_H)$. 
Note that we connect our EoSs to the integral pQCD constraint at the TOV point. The actual location of the TOV point, i.e. $(\mu_c^{\mathrm{TOV}},n_c^\tov)$ changes from one EoS to another, as shown in Fig.~\ref{fig:n_mu}. 
Beyond $n_c^\tov$ the NS branch is unstable and not observable. Furthermore, in this work, we are interested in determining the constraining power of pQCD at densities relevant for NSs. This makes $n_c^\tov$ the natural density at which the connection to pQCD should be performed. Matching at a larger density could induce model dependencies based on how the extrapolation to the larger density is performed.
The impact of the matching density is studied in more detail later in this article.

The contours defined by Eqs.~\eqref{eq:n_max}-\eqref{eq:n_min} are shown in Fig.~\ref{fig:n_mu} in red. Note that the region encapsulated by the red contours represent a necessary but not sufficient condition to fulfill the pQCD constraint, since every EoS has to satisfy the additional criteria $p_\textrm{min}(\mu,n) < p(\mu,n) < p_\textrm{max}(\mu,n)$ where
\begin{equation}
    p_\textrm{min}(\mu,n) = p_L + \frac{\mu^2 - \mu_L^2}{2\mu} n_\textrm{min}(\mu) 
    \label{eq:p_min}
\end{equation}
and 
\begin{equation}
  p_\textrm{max}(\mu,n) =
  \begin{cases}
   p_L + \frac{\mu^2 - \mu_L^2}{2\mu} n  & n < n_c(\mu) \\
    p_H - \frac{\mu_H^2 - \mu^2}{2\mu} n  & n > n_c(\mu)
  \end{cases}
  \label{eq:p_max}
\end{equation}
where $n_c(\mu) = n_\textrm{max}(\mu_L)\mu/\mu_L$.
In the top panel of Fig~\ref{fig:n_mu}, we have considered $\mu_{\textrm{pQCD}}=2.6$~GeV as suggested in Ref.~\cite{Gorda:2021znl}, and we explore the sensitivity of the results to this choice in the bottom panel for $\mu_{\textrm{pQCD}}=2.4$~GeV. 
At fixed $\mu_{\textrm{pQCD}}$, the uncertainties in the pQCD EoS can be estimated by varying the renormalization scale parameter $X$ as in Ref.~\cite{Komoltsev:2021jzg}. 
In Fig.~\ref{fig:n_mu}, we show  results for $X=[1,2,4]$.

\begin{figure*}
    \centering
    \includegraphics[height=0.27\textwidth,width = 0.32\textwidth]{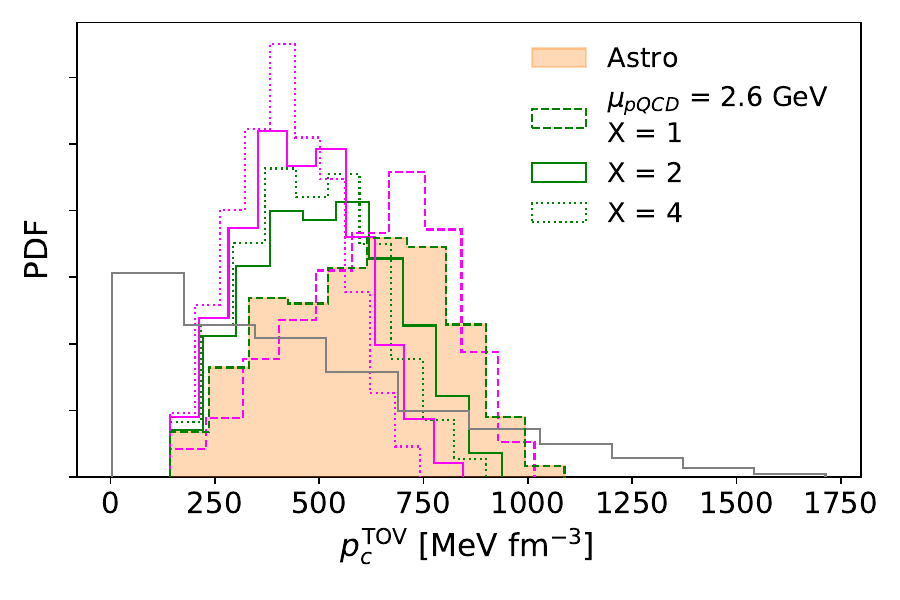}
    \includegraphics[height=0.27\textwidth,width = 0.32\textwidth]{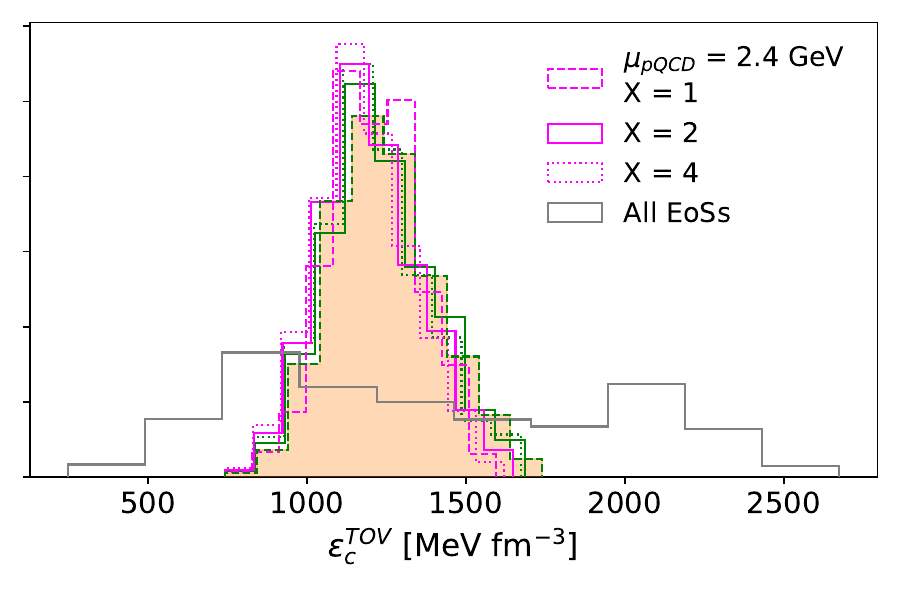}
    \includegraphics[height=0.27\textwidth,width = 0.32\textwidth]{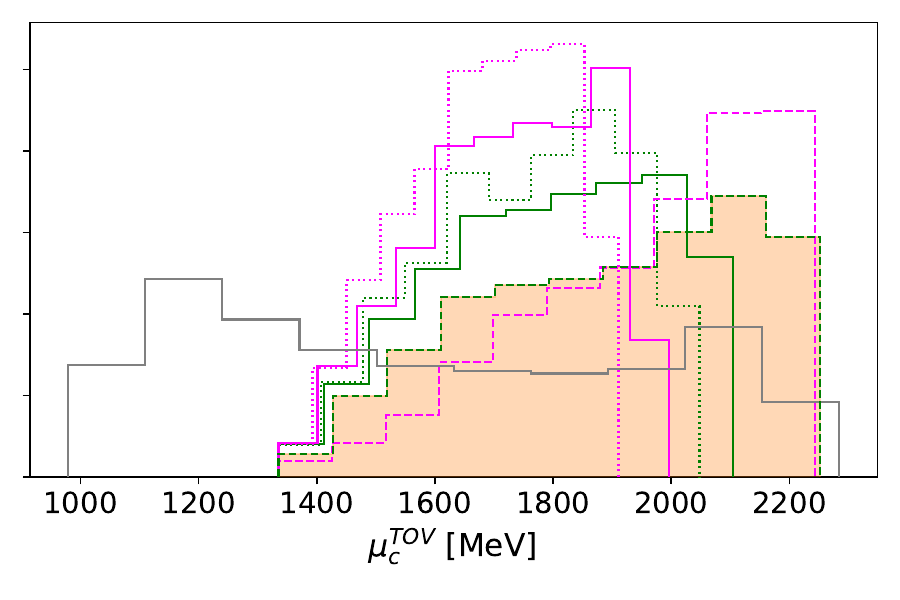}
    \caption{The PDFs for the central pressure (left), the central energy density (middle) and central chemical potential (right) in maximally massive NS. The shaded orange PDF is obtained by imposing astrophysical constraints only, whereas the green and magenta PDFs impose pQCD constraints on top of the astrophysical constraints.
    }
    \label{fig:p_tov}
\end{figure*}

\textit{Results for the EoS -}
Several EoS in our set are consistent with astrophysical NS observations but inconsistent with pQCD constraints for $X=2$ and $X=4$, see Fig.~\ref{fig:n_mu}. 
These EoS are too stiff below $n_c^\tov$, leading to a fast rise of the chemical potential with the number density. The chemical potential then becomes too large to be connected to the pQCD limit. 
This violation of the pQCD constraint is even more pronounced for $\mu_{\mathrm{pQCD}} = 2.4$~GeV, leading to an interesting interplay between the pQCD constraining power and astrophysical observations. Current astrophysical observations of NS are consistent with stiffer EoS at high densities to account for the existence of $2\Msol$ NS~\cite{Cromartie:2019} and their possibly large radii suggested by NICER~\cite{Miller:2021qha,Riley:2021pdl}. 
Hence, these data require a rapid increase of the pressure as function of the energy density $\epsilon$, which in turn, implies large values for the speed of sound $c_s^2=\partial p/\partial\epsilon$. 
Because the speed of sound can also be expressed as $c_s^2=(n \partial \mu)/(\mu \partial n)$, this implies that the chemical potential rises rapidly with the number density. In Fig.~\ref{fig:n_mu}, we see that for some EoS chemical potentials as large as $\mu \approx 2.2$~GeV can be reached in the center of maximally massive NS, which is comparable to $\mu_{\mathrm{pQCD}}$. 
If the number density for this configuration is too low, the asymptotic pQCD limit cannot be reached in a thermodynamically consistent manner and, therefore, such EoS are ruled out. This clarifies how pQCD impacts dense matter at NS densities even though pQCD itself is valid only at much larger densities: while NS never explore densities close to $n_{\mathrm{pQCD}}$, they might explore chemical potentials close to $\mu_{\mathrm{pQCD}}$.

Note that, for the Probability Density Functions (PDF) shown in Fig.~\ref{fig:p_tov}, the pQCD constraints are imposed ``on top of'' the astrophysical constraints. We see that the dashed lines corresponding to $X=1$, with $\mu_{\textrm{pQCD}}=2.6$~GeV, coincide with the astro-only PDF, showing that pQCD has no impact in this case. 
For $p_c^{\mathrm{TOV}}$, pQCD constraints with $X=2$ and $X=4$ reduce the maximal pressure explored in NS, with the effect being more pronounced for larger values of $X$ and lower values of $\mu_{\mathrm{pQCD}}$. 
Interestingly the case $X=1$ with $\mu_{\textrm{pQCD}}=2.4$~GeV clearly impacts the PDF for $p_c^{\mathrm{TOV}}$ and $\mu_c^{\mathrm{TOV}}$ by excluding certain soft EoS.

\begin{figure*}[t]
    \centering
    \includegraphics[height=0.27\textwidth,width = 0.32\textwidth]{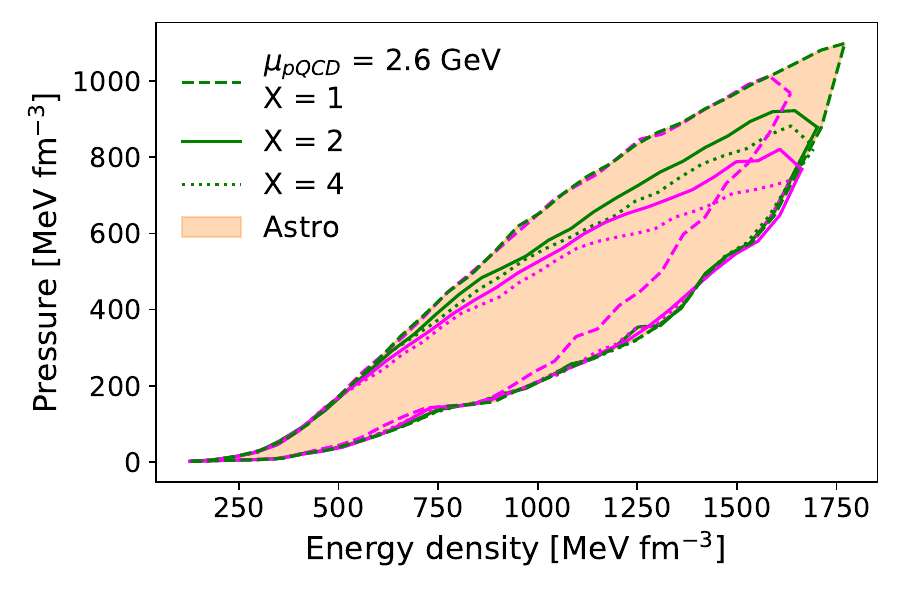}
    \includegraphics[height=0.27\textwidth,width = 0.32\textwidth]{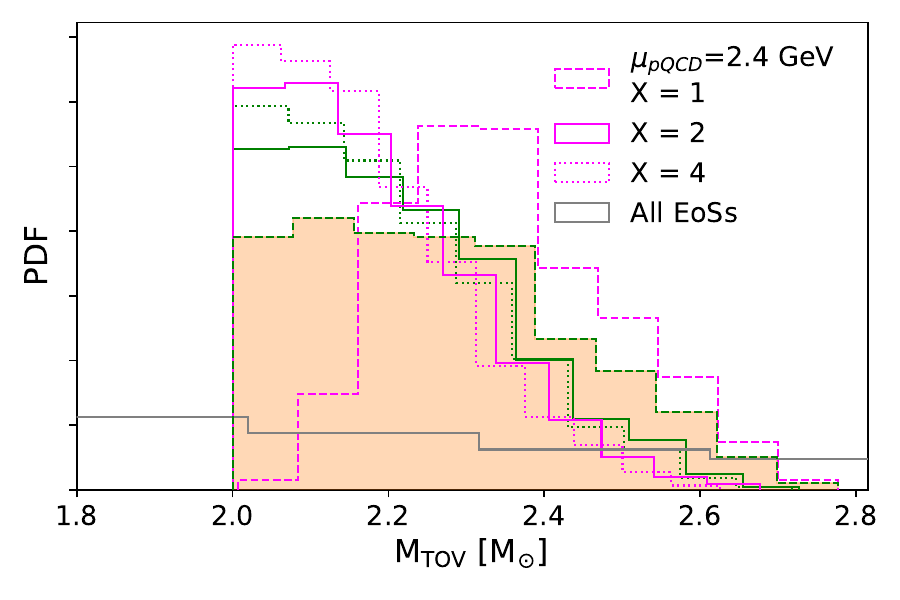}
    \includegraphics[height=0.27\textwidth,width = 0.32\textwidth]{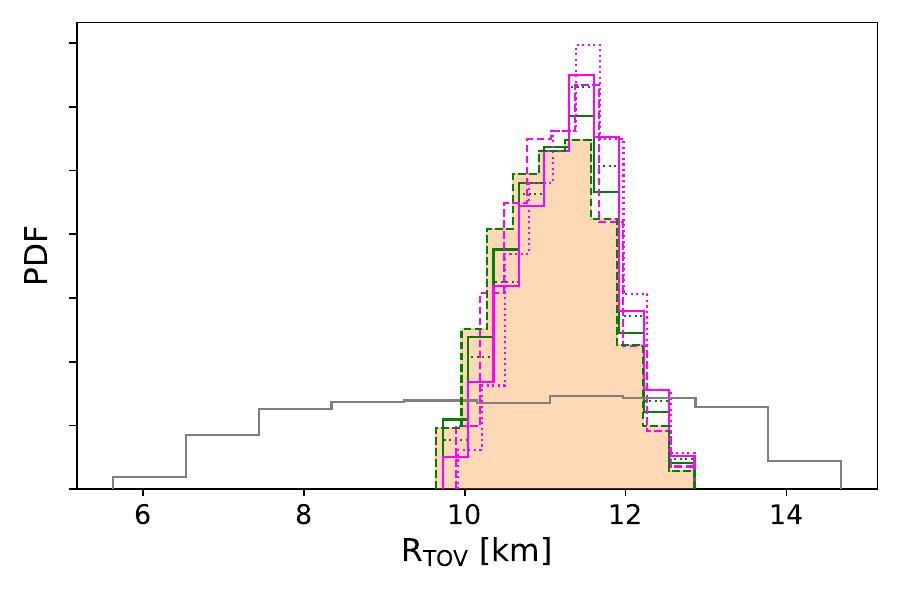}
    \caption{Left panel: pressure as a function of the energy density where the upper and lower edges of the contours represent the envelop over all the EoS allowed in the corresponding set.
    PDFs over the maximum NS mass (middle panel) and its corresponding radius (right panel) are also shown. }
    \label{fig:M_tov_dist}
\end{figure*}

In the left panel of Fig.~\ref{fig:M_tov_dist}, we show envelopes for the pressure as function of the energy density around all the EoS allowed in a certain set. The bands are terminated at the TOV limit. For the cases $X=2$ and $X=4$, we find that incorporating pQCD constraints reduces the allowed region in the $p - \epsilon$ plane by lowering the maximum allowed pressures. Again, the pQCD constraints do not reduce the number of EoS if $X=1$ and $\mu_{\textrm{pQCD}}=2.6$~GeV. 
In general, for $\mu_{\textrm{pQCD}}=2.6$~GeV, we have found that the minimal value of $X$ at which pQCD becomes constraining is around $X \approx 1.3$.
Note the impact of pQCD if $X=1$ and $\mu_{\textrm{pQCD}}=2.4$~GeV as previously discussed.

\textit{Results for the NS masses and radii -}
In the middle and right panels of Fig.~\ref{fig:M_tov_dist}, we show the PDFs over the maximal NS mass M$_\tov$ and the radius of the corresponding NS, R$_\tov$. For M$_\tov$, the effects of pQCD are most significant when the case $X=1$ and $\mu_{\textrm{pQCD}}=2.4$~GeV is considered.
For $X=2$ and $X=4$, we find that pQCD constraints slightly shift the distributions for M$_\tov$ to lower values when added on top of astrophysical constraints but the corresponding PDF over R$_\tov$ show no significant change. 
These results indicate that present pQCD calculations do not impact the masses and radii of observable NS, but they are on the brink of becoming constraining. 

\begin{figure}
    \centering
    \includegraphics[scale=0.49]{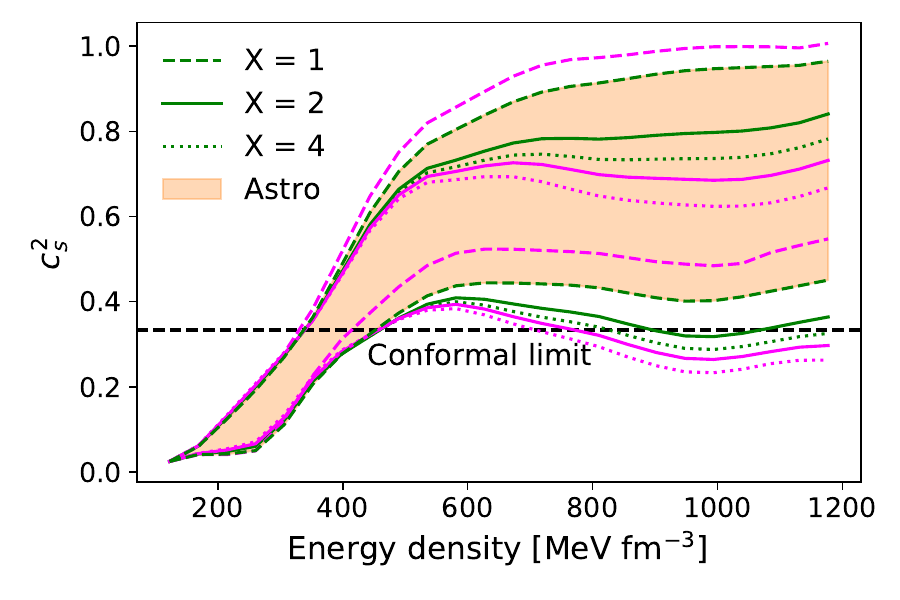}
    \includegraphics[scale=0.49]{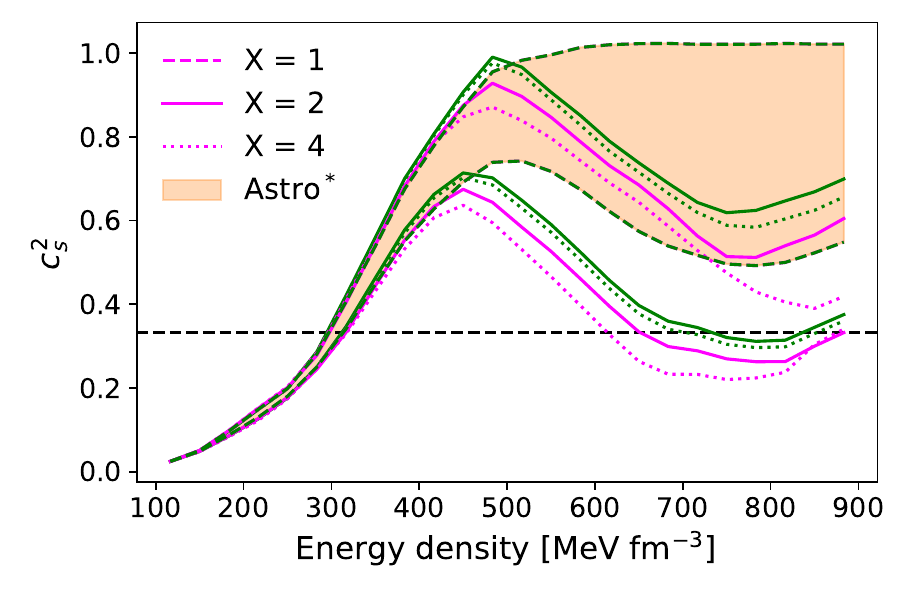}
    \caption{The speed of sound as function of the energy density. The upper and lower edges represent a $2 \sigma$ uncertainty where
    $\sigma$ is the sample standard deviation of our EoS ensemble. 
    Top: Constraints on the speed of sound for NS observational data to date for a variation of the pQCD renormalization scale $X$. 
    Bottom: Constraints on the speed of sound for NS observational data when additionally imposing $R_{2.0} > 13$~km (Astro$^*$). Note that the $X=1$ magenta dashed lines are not visible since they overlap with the $X=1$ green case.}
    \label{fig:c2_astro}
\end{figure}

\textit{Impact of potentially more constraining future measurements -}
Thus far, we have addressed the interplay of astrophysical observations, requiring a stiffening of the EoS, and pQCD calculations, requiring a softening of the EoS. 
Now, we investigate how an improved future measurement of the radius $R_{2.0}$ of a two solar mass NS would influence our findings. In the top panel of Fig.~\ref{fig:c2_astro}, we show constraints on the sound speed including observational NS data to date. Additionally imposing constraints from pQCD slightly lowers the average sound speed when larger values of $X$ are considered, but we find a significant overlap of EoS ranges with or without pQCD constraint. 
In the bottom panel of Fig.~\ref{fig:c2_astro}, we now impose the additional constraint $R_{2.0} > 13$~km, which is a possible future scenario given the NICER measurement of PSR J0740+6620~\cite{Miller:2021qha,Riley:2021pdl}. 
We find that in this case, adding pQCD constraints on top of astrophysical data shows a more significant impact, leading to the formation of a pronounced peak in the sound speed for $X=2$ and $X=4$: 
At low energy densities, the sound speed rises rapidly due to the imposed constraint $R_{2.0} > 13$~km; at $\epsilon \gtrsim 500 \MeV \fmiq$, the speed of sound plateaus if only astrophysical data are considered, while it decreases significantly if pQCD (at larger $X$ values) is added. 
Such non-monotonous behaviour of the sound speed is expected to be indicative of the appearance of exotic, non-nucleonic degrees of freedom~\cite{Somasundaram:2021clp,Tan:2020ics,Tan:2021nat}, such as quarkyonic matter as suggested in Ref.~\cite{McLerran:2019}.

\textit{Impact of changing the matching density -}
Recently, the impact of pQCD on the inference of the NS EoS has been studied by Ref.~\cite{Gorda:2022jvk}, using an approach similar to ours. However, their results seem to indicate a larger impact of pQCD on the NS EOS. There are three main differences in the two approaches:
\begin{itemize}
    \item We match to the pQCD constraint at $n_c^\tov$ while Ref.~\cite{Gorda:2022jvk} matches at $10n_{\rm sat}$,
    \item Ref.~\cite{Gorda:2022jvk} uses Gaussian Processes (GPs) to generate EoS models while we use a speed-of-sound model,
    \item and Ref.~\cite{Gorda:2022jvk} uses a probabilistic approach while we study envelopes. 
\end{itemize}
We argue that point 1 is the most important factor.

\begin{figure}
    \centering
    \includegraphics[scale=0.55]{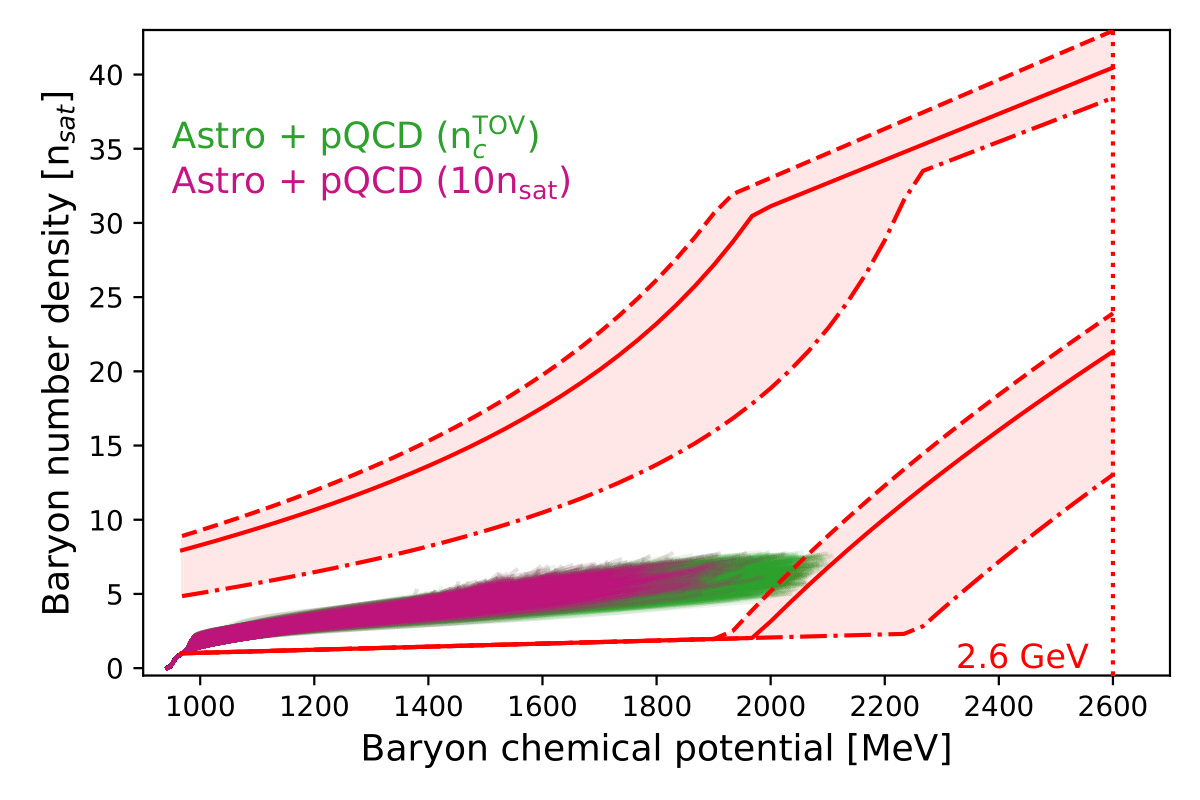}
    \caption{The baryon number density as function of the chemical potential, similar to the top panel of Fig.~\ref{fig:n_mu}. Additionally, we show results for a matching density of  $10 n_{\rm sat}$ in purple. Each EoS is plotted up to the TOV limit.}
    \label{fig:n_mu_10nsat}
\end{figure}

\begin{figure}[t]
    \centering
    \includegraphics[scale=0.55]{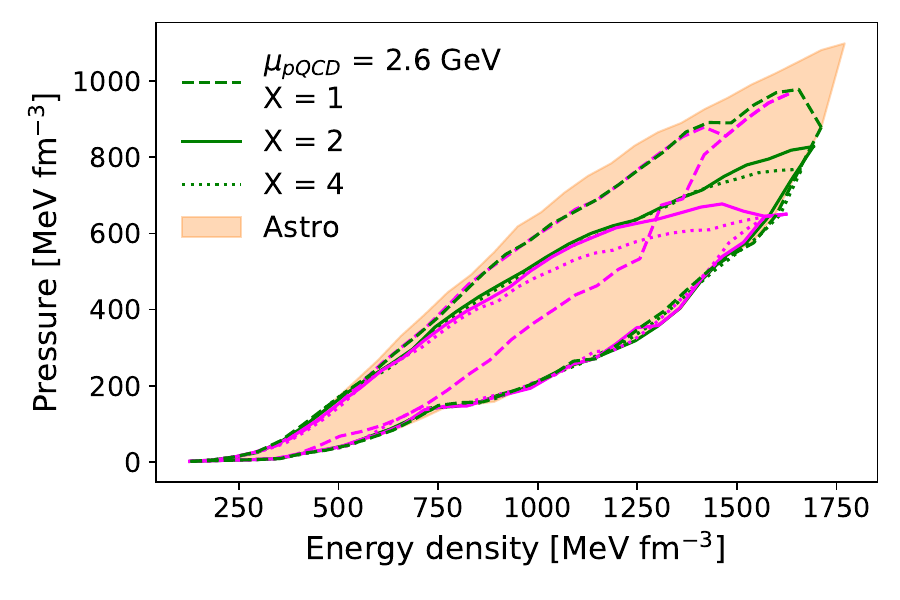}
    \caption{Pressure as a function of the energy density similar to Fig.~\ref{fig:M_tov_dist} but for a matching density of $10 n_{\rm sat}$. As before, the upper and lower edges of the contours represent the envelop over all the EoS allowed in the corresponding set.}
    \label{fig:EOS_10nsat}
\end{figure}

Regarding the point 3, when studying the impact of pQCD on the nuclear EOS it is most conservative to study envelopes as they account for all possible EoS behavior. 
Probabilistic treatments, on the other, could smear out the existence of phase transitions because these are more fine-tuned EOS models. 

Regarding the second point, while different EOS parametrizations might impact results in a probabilistic framework (see, e.g., Ref.~\cite{Legred:2022pyp}),
EoS envelopes are less sensitive to this choice and the speed-of-sound parametrization can equally well capture extreme behavior compared to GPs. 
Furthermore, previous EoS inferences using similar astrophysical and EoS constraints agree well, independent of using GPs or a speed-of-sound parametrizations~\cite{Dietrich:2020lps,Essick:2020flb}.
Hence, we do not expect this choice to influence our results drastically.

We, therefore, expect that the difference between our results and those of Ref.~\cite{Gorda:2022jvk} arise primarily due to point 1. To investigate this, we repeated our study using a matching density of $10n_{\rm sat}$; see Figs.~\ref{fig:n_mu_10nsat} and~\ref{fig:EOS_10nsat}.
In Fig.~\ref{fig:n_mu_10nsat}, the result obtained using a matching density of  $10n_{\rm sat}$ is compared with that obtained using $n_c^\tov$, whereas in Fig.~\ref{fig:EOS_10nsat}, we show results only for the $10n_{\rm sat}$ case. The EoSs that are very stiff inside NSs have a larger probability of remaining somewhat stiff beyond $n_c^\tov$ than for the EOS to undergo a phase transition,
leading to a likely violation of the pQCD limit if the matching is performed at larger densities.
We find that changing this matching density to $10n_{\rm sat}$ leads to an exclusion of stiff EoS, in complete agreement with the findings of Ref.~\cite{Gorda:2022jvk}.

Because the density where pQCD integrals are connected plays an important role for making claims about the constraining power of pQCD, constraints from pQCD are very sensitive to model assumptions. 
This emphasizes our findings that pQCD is currently not constraining but might be so in future, when uncertainties are reduced such that simple model choices do not matter as much.

\textit{Conclusions -}
We have systematically studied the effects of incorporating pQCD calculations when analyzing the EoS of NS. 
Using a model-independent approach to the EoS, we concluded that pQCD does not significantly constrain the EoS, if it is imposed on top of current observational constraints on NS taking their uncertainties into account.
However, the calculations are at the brink of being constraining for NS: improved pQCD constraints or new astrophysical data preferring stiff EoS will reveal the potential of pQCD for EoS selection. 
We found an interesting interplay of astrophysical observation and pQCD calculations: While certain astrophysical observations of NS are consistent with stiff EoS, supporting large NS masses and radii, such EoS are disfavored upon imposing pQCD constraints. 

These findings have important implications for the study of dense matter. 
In particular, while an analysis of future astrophysical data alone might be inconclusive with respect to the existence of exotic matter in neutron-star cores, the combined analysis of astrophysical and pQCD data might help answering this question.
Therefore, future work on improving pQCD constraints is crucial for the studies of dense neutron-star matter.


We thank A. Kurkela, T. Gorda and O. Komoltsev for helpful feedback on the manuscript.
R.S. is supported by the PHAST doctoral school (ED52) of \textsl{Universit\'e de Lyon}. R.S. and J.M. are both supported by CNRS grant PICS-08294 VIPER (Nuclear Physics for Violent Phenomena in the Universe), the CNRS IEA-303083 BEOS project, the CNRS/IN2P3 NewMAC project, and benefit from PHAROS COST Action MP16214 as well as the LABEX Lyon Institute of Origins (ANR-10-LABX-0066) of the \textsl{Universit\'e Claude Bernard Lyon-1}.
The work of I.T. was supported by the U.S. Department of Energy, Office of Science, Office of Nuclear Physics, under contract No.~DE-AC52-06NA25396, by the Laboratory Directed Research and Development program of Los Alamos National Laboratory under project number 20220658ER, and by the U.S. Department of Energy, Office of Science, Office of Advanced Scientific Computing Research, Scientific Discovery through Advanced Computing (SciDAC) NUCLEI program.

\bibliography{biblio}
\end{document}